\documentclass[12pt]{article}
\usepackage{a4}
\usepackage{amsfonts,amssymb,amsmath,amsthm}
\usepackage{graphicx}

\newcommand{\is}{{\Sigma \hspace{-1.1em} \int}}

\begin{document}
\title{On space-time noncommutative theories at finite temperature}
\author{A.~V.~Strelchenko\thanks{E.mail:
\texttt{alexstrelch(at)yahoo.com}}\\{\em Dnepropetrovsk National University,
49050 Dnepropetrovsk, Ukraine}\\[5pt]
D.~V.~Vassilevich\thanks{On leave from V.~A.~Fock Institute of
Physics, St.~Petersburg University, Russia.
E.mail:\ {\texttt{dmitry(at)dfn.if.usp.br}}}\\
{\it Instituto de F\'isica, Universidade de S\~ao Paulo,}\\ {\it
Caixa Postal 66318 CEP 05315-970, S\~ao Paulo, S.P., Brazil}}

\maketitle
\begin{abstract}
We analyze renormalization and the high temperature expansion
of the one-loop effective action of the space-time noncommutative
$\phi^4$ theory by using the zeta function regularization in the
imaginary time formalism (i.e., on $S^1\times \mathbb{R}^3$).
Interestingly enough, there are no mixed (non-planar) contributions
to the counterterms as well as to the power-law high temperature
asymptotics. We also study the Wick rotation and formulate assumptions
under which the real and imaginary time formalisms are equivalent.
\end{abstract}
\section{Introduction}
Since a number of review papers have been published recently (see
\cite{NCrev}) it is not necessary to repeat here the motivations
for studying noncommutative (NC) field theories. Most of the
previous previous works, see e.g.
\cite{Arcioni:1999hw,Fischler:2000fv,Landsteiner:2000bw,Brandt:2006bf} and
references therein, on finite temperature NC theories analyzed the
case of space-space noncommutativity (with very few exceptions
\cite{Brandt:2003je,Brandt:2007yg}). Indeed, the case of space-time
noncommutativity is most problematic because of the difficulties
with unitarity and causality which were discovered some years ago
\cite{unicas,GoMe,AGBZ}. These difficulties have not been
completely resolved up to now. Space-time NC theories with compact
dimensions exhibit an interesting phenomenon of discreteness of
time \cite{distime}.

The main purpose of this paper is to develop certain aspects of
the Euclidean space formalism in space-time NC theories, including
the renormalization, the transition from real to imaginary time,
and the high temperature asymptotics.

We shall start our work with analyzing the one-loop divergences in
the Euclidean NC $\phi^4$ on $S^1\times \mathbb{R}^3$ to make sure
that the theory which will be discussed later does exist at least
at the leading order of the loop expansion. We shall use the
zeta-function regularization \cite{zetareg,Elizalde:1994gf} and
the heat kernel technique
\cite{KirstenBook,NewGilkey,Vassilevich:2003xt}. In the context of
an NC field theory the heat kernel expansion was first obtained
for the operators which contain only left or only right Moyal
multiplications \cite{Vassilevich:2003yz,Gayral:2004ww}. Such
operators were, however, insufficient to deal with some physical
applications, like, for example, the $\phi^4$ theory. The heat
kernel expansion for generalized Laplacians containing {\em both}
left and right Moyal multiplications was constructed in
\cite{Vassilevich:2005vk} on the Moyal plane and in
\cite{Gayral:2006vd} on the NC torus. Non-minimal operators were
considered in \cite{Strelchenko:2006za}. We would also like to
mention the calculations \cite{GrosseHK} of the heat kernel
expansion in the NC $\phi^4$ model modified by an oscillator-type
potential.

To avoid unnecessary technical complications we shall study
exclusively the case of pure space-time noncommutativity, i.e., we
put to zero the NC parameter with both indices in the spatial
directions, $\theta^{jk}=0$. We shall calculate the heat kernel
coefficients $a_n$ with $n\le 4$. It will appear that the
coefficients $a_2$ and $a_4$ look very similar to the commutative
theory, but $a_3$ is given by a complicated non-local expression.
Fortunately, odd-numbered heat kernel coefficients do not
contribute to one-loop divergences at four dimensions in the
zeta-function regularization. The model will turn out to be
one-loop renormalizable with temperature-independent counterterms.

Of course, we do not expect this model to be renormalizable at all
loops. There are well-known problems related to the so-called
UV/IR mixing \cite{UVIR} which should also be present in our case
(though, maybe, in a relatively mild form since one of the NC
directions is compact). To make the finite temperature NC $\phi^4$
renormalizable to all orders one should probably make it duality
covariant \cite{GW} or use a bifermionic NC parameter
\cite{biferm}.

An approach to finite temperature theories on static backgrounds
based on the zeta-function regularization was developed long ago
by Dowker and Kennedy \cite{DowKen}. In particular, they
established relations between spectral functions of a
3-dimensional operator which defines the spectrum of fluctuations
and the high temperature asymptotics of the free energy. In our
case, due to the presence of the space-time noncommutativity, such
3-dimensional operator becomes frequency dependent even on static
backgrounds. Therefore, eigenfrequencies of fluctuations are
defined by a sort of a non-linear spectral problem. Fortunately
for us, a technique which allows to work with finite temperature
characteristics of the theories leading to non-linear spectral
problems has been developed relatively recently in the papers
\cite{nlsp}. These papers were dealing with the thermodynamics of
stationary but non-static space-times, but, after some
modifications, the approach of \cite{nlsp} can be made suitable
for space-time noncommutative theories as well. By making use of
these methods we shall construct the spectral density of states in
the real-time formalism and express it through the heat kernel of
a frequency dependent operator in three dimensions. Then, by using
this spectral density, we shall demonstrate that the Wick rotation
of the Euclidean free energy gives the canonical free energy. To
come to this conclusion we shall need two assumptions. First of
all, we shall have to assume that the spectral density behaves
"nicely" as a function of complex frequencies. Although this
assumption is very hard to justify rigorously, we shall argue that
the behavior of the spectral density must not be worse than in
the commutative case, and we shall also suggest a consistency
check based on the high temperature asymptotics. There is no
canonical Hamiltonian in the space-time NC theories. Therefore, we
have to assume that the eigenfrequncies of quantum fluctuations
can replace one-particle energies in thermal distributions. This
assumptions cannot be derived from the first principles of
quantization basing on the present knowledge on the subject, but
we can turn the problem around: the very fact that the Wick
rotation of the Euclidean free energy leads to a thermal
distribution over the eigenfrequences supports (a rather natural)
guess that the eigenfrequences are the energies of one-particle
excitations. Let us stress, that the calculations we shall perform
in the Euclidean space do not depend on the assumptions described
above.

We shall also use the heat kernel methods to calculate the high temperature
asymptotics of the Euclidean effective action assuming that the background
field is static. As in the case of
the counterterms, there are non-planar contributions. The asymptotic
expansion does not depend on the NC parameter (provided it is
non-zero) and looks very similar to the commutative case.

This paper is organized as follows. In the next section we consider
one-loop renormalization of NC $\phi^4$ on $S^1\times \mathbb{R}^3$.
Sec.\ \ref{sec-IR} is devoted to the Wick rotation. High temperature
asymptotics of the effective action are calculated in sec.\ \ref{sec-hT}.
Some concluding remarks are contained in sec.\ \ref{sec-con}.

\section{Noncommutative quantum field theory on $S^1\times \mathbb{R}^3$}
\subsection{Basic definitions and notations}
Let us consider a scalar $\phi^{4}$ model on NC $S^1 \times
\mathbb{R}^3$. The scalar field is periodic with respect to the
compact coordinate. We use the notations $(x^\mu)
=(\bar{x},x^4)=(x^i,x^4)$, where $x^4$ is a coordinate on $S^1$,
$0 \leq \tau \leq \beta$. Similarly for the Fourier momenta we use
$k=(\bar{k},k_4)$, $k_{4}=\frac{2\pi l}{\beta}$, $l \in
\mathbb{Z}$.

The action reads
\begin{equation}\label{s1}
 S=\frac{1}{2} \int_{0}^{\beta} dx^4~\int_{\mathbb{R}^3}
 d^3\bar x~\left((\partial_{\mu}\phi)^2
 +m^{2}\phi^{2}+\frac{g}{12}~\phi_{\star}^{4}\right),
\end{equation}
where the $\phi_{\star}^{4}=\phi \star \phi \star \phi \star
\phi$. Star denotes the Moyal product
\begin{equation}
f_1\star f_2 (x)=\exp \left( \frac \imath 2 \theta^{\mu\nu} \partial_\mu^x
\partial_\nu^y \right) f_1(x) f_2(y)\vert_{y=x} .\label{Mprod}
\end{equation}
To simplify the setup we assume that $\theta^{ij}=0$, but some of
$\theta^{4j}\ne 0$, i.e. we have an Euclidean space-time
noncommutativity.

We wish to investigate quantum corrections to (\ref{s1}) by means
of the background field method. To this end one has to split the
field $\phi$ into a classical background  field $\varphi$
 and quantum
fluctuations, $\phi=\varphi + \delta \varphi$. The one-loop
contribution to the effective  action is defined by the  part of
(\ref{s1}) which is quadratic in quantum fluctuations:
\begin{equation}\label{s2}
 S[\varphi, \delta \varphi]=\frac{1}{2}\int_{0}^{\beta} dx^4\ \int_{\mathbb{R}^3}
 d^3 \bar x~\delta
\varphi (D+m^2) \delta \varphi,
\end{equation}
where the operator $D$ is of the form (cf.
\cite{Gayral:2004cu,Vassilevich:2005vk}) :
\begin{equation}\label{one}
  D=-(\partial_\mu \partial^\mu +E),
\end{equation}
with
\begin{equation}
E=-\frac{g}{6}\left(L(\varphi \star \varphi) +R(\varphi \star
\varphi)+L(\varphi) R(\varphi)\right). \label{Ephi4}
\end{equation}

The one-loop effective action can be formally written as
\begin{equation}\label{s3}
 W=\frac{1}{2}\ln\det (D+m^2).
\end{equation}
This equation still has to be regularized. To make use of the
zeta-function regularization we have to define the heat
kernel\footnote{A better name used in mathematics for this object
is the heat trace, but here we use the terminology more common in
physics.}
\begin{equation}
K(t,D)={\rm Tr} \left( e^{-tD}-e^{tD_0}\right) \label{heattr}
\end{equation}
and the zeta function
\begin{equation}
\zeta (s,D+m^2) ={\rm Tr}\left( (D+m^2)^{-s} - (D_0+m^2)^{-s}
\right). \label{defzeta}
\end{equation}
Here ${\rm Tr}$ is the $L_2$ trace. In both cases we subtracted
the parts corresponding to free fields with $D_0=-\partial_\mu
\partial^\mu$ to avoid volume divergences.

The regularized one-loop effective action is defined as
\begin{equation}
W_s=-\frac 12 \mu^{2s} \int_0^\infty \frac{dt}{t^{1-s}} e^{-tm^2}
K(t,D)=-\frac 12 \mu^{2s} \Gamma(s) \zeta (s,D+m^2),\label{zereg}
\end{equation}
where $s$ is a regularization parameter, $\mu$ is a constant of
the dimension of mass introduced to keep proper dimension of the
effective action. The regularization is removed in the limit $s\to
0$. At $s=0$ the gamma-function has a pole, so that near $s=0$
\begin{equation}
W_s=-\frac 12 \left( \frac 1s -\gamma_E +\ln \mu^2 \right)\zeta
(0,D+m^2) -\frac 12 \zeta' (0,D+m^2),\label{near0}
\end{equation}
where $\gamma_E$ is the Euler constant.

Let us assume that there is an asymptotic expansion of the heat
kernel as $t\to +0$
\begin{equation}
K(t,D)=\sum_{n=1}^\infty t^{(n-4)/2} a_n(D) \,.\label{asymptotex}
\end{equation}
Such an expansion exists usually (but not always) in the
commutative case. On NC $S^1\times \mathbb{R}^3$ the existence of
(\ref{asymptotex}) will be demonstrated in sec.\ \ref{sec-hk}. For
a Laplace type operator on a commutative manifold all odd-numbered
heat kernel coefficients $a_{2k-1}$ vanish. (They are typical
boundary effects). As we shall see below, on NC $S^1\times
\mathbb{R}^3$ the coefficient $a_3\ne 0$. The coefficient $a_0$
disappears due to the subtraction of the free-space contribution
in (\ref{heattr}).

The pole part of $W_s$ can be now expressed through the heat kernel
coefficients.
\begin{equation}
\zeta(0,D+m^2) = -m^2a_2(D) + a_4(D).\label{zeta0}
\end{equation}
Note, that odd-numbered heat kernel coefficients $a_{2p-1}(D)$ do
note contribute to the divergences of $W_s$.
\subsection{Heat kernel expansion on $S^1\times
\mathbb{R}^3$}\label{sec-hk}

 Let us consider the operator
\begin{equation}
D=-(\partial_\mu^2 + E),\qquad
E=L(l_1)+R(r_1)+L(l_2)R(r_2)\label{opD}
\end{equation}
on $S^1\times \mathbb{R}^3$. This operator is slightly more
general than the one in (\ref{one}). The potential term
(\ref{Ephi4}) is reproduced by the choice
\begin{equation}
l_1=r_1=-\frac g6 \varphi \star \varphi
\,,\quad l_2=-r_2=\sqrt{\frac g6} \varphi \,.\label{choi}
\end{equation}
We are interested in the asymptotics of the heat trace
(\ref{heattr}) as $t\to +0$. To calculate the trace we, as usual,
sandwich the operator between two normalized plane
waves\footnote{Although we are working with a real field, it is
more convenient to use complex plane waves instead of real
functions $\sin (kx)$ and $\cos (kx)$. For a complex field we
would have a coefficient of $1$ instead of $1/2$ on the right hand
side of (\ref{s3}). Since $D$ with (\ref{choi}) is real, this is
the only difference.} , and integrate over the momenta and over
the manifold $\mathcal{M}=S^1\times \mathbb{R}^3$.
\begin{equation}
K(t;D)=\frac 1{\beta (2\pi)^3}\, \is dk\, \int_{\mathcal M} d^4x\,
e^{-\imath kx}
\left( e^{-tD}  -e^{tD_0}\right)e^{\imath kx} ,\label{ptrace}
\end{equation}
where we introduced the notation
\begin{equation}
\is dk\equiv \sum_{k_4} \int d^3\bar k \label{is}
\end{equation}
with $k_4=2\pi n/\beta$, $n\in \mathbb{Z}$. To evaluate the asymptotic
expansion of (\ref{ptrace}) at $t\to +0$ one has to extract the factor
$e^{-tk^2}$.
\begin{eqnarray}
K(t,D)&=&\frac 1{\beta (2\pi)^3}\int d^4x \is dk e^{-tk^2}\nonumber \\
&&\times \langle \exp \left( t\left( (\partial -ik)^2
+2i k^\mu (\partial_\mu -ik_\mu)
+E\right)\right)-1\rangle_k ,\label{expk2}
\end{eqnarray}
where we defined
\begin{equation}
\langle F \rangle_k \equiv e^{-\imath kx}\star F e^{\imath kx} \label{angle}
\end{equation}
for any operator $F$. Next one has to expand the exponential
in (\ref{expk2}) in a power series in $E$ and $(\partial -\imath k)$.
As we shall see below, only a finite number of terms in this
expansion contribute to
any finite order of $t$ in the $t\to +0$ asymptotic expansion of
the heat kernel. We push all $(\partial -\imath k)$ to the right until
they hit $e^{\imath kx}$ and disappear.
\begin{eqnarray}
K(t,D)&=&\frac 1{\beta (2\pi)^3}\int d^4x \is dk e^{-tk^2}
\left\langle tE + \frac {t^2}2 ([\partial_\mu,[\partial_\mu,E]]
+E^2 +2\imath k^\mu [\partial_\mu,E])\right.
\nonumber \\
&&\qquad \left. -\frac{4t^3}6 k^\mu k^\nu [\partial_\mu,[\partial_\nu,E]]
+\dots \right\rangle_k \,.\label{expk3}
\end{eqnarray}
We kept in this equation all the terms which may contribute to
$a_n$ with $n\le 4$. In the commutative case all total derivatives
as well as all term linear in $k$ vanish. In the NC case this is
less obvious because of the non-locality, so that we kept also
such terms. The commutator of $\partial_\mu$ with $E$ is a
multiplication operator, e.g., $[\partial_\mu,L(l)]=L(\partial_\mu
l)$, $[\partial_\mu,L(l)R(r)]=L(\partial_\mu
l)R(r)+L(l)R(\partial_\mu r)$. Therefore, eq.\ (\ref{expk3})
contains multiplication operators of two different sorts: the ones
with only left or only right Moyal multiplications, and the ones
containing products of left and right Moyal multiplications. The
terms of different sorts will be treated
differently\footnote{Formally $R(r)=L(1)R(r)$, but a constant
function does not belong to $C^\infty(S^1\times \mathbb{R}^3)$
since it does not satisfy the fall-off conditions. Consequently,
the two sorts of the term discussed above indeed lead to quite
different asymptotics at $t\to +0$. }.

The terms with one type of the multiplications are easy. We shall
call such terms planar borrowing the terminology from the approach
based on Feynman diagrams. They can be evaluated in the same way
as in \cite{Vassilevich:2003yz,Gayral:2004ww}. Because of the
identities
\begin{equation}
\int d^4x \langle R(r) \rangle_k = \int d^4x\, r(x)\,,\qquad \int
d^4x \langle L(l) \rangle_k = \int d^4x\, l(x)\label{Rrr}
\end{equation}
only the $E$ and $E^2$ terms in (\ref{expk3}) contribute. It remains
then to evaluate the asymptotics of the integral
\begin{equation}
\frac 1{\beta (2\pi)^3} \is dk\, e^{-tk^2}=(4\pi t)^{-2} +
\mbox{e.s.t.},\label{askin}
\end{equation}
where e.s.t.\ denotes exponentially small terms, to obtain
\begin{eqnarray}
&&a_2^{\rm planar}(D)=(4\pi)^{-2} \int d^4x (l_1+r_1),\label{a2plan}\\
&&a_4^{\rm planar}(D)=(4\pi)^{-2} \int d^4x \frac 12 (l_1^2+r_1^2).
\label{a4plan}
\end{eqnarray}

Non-planar (mixed) contributions require considerably more work.
The typical term reads
\begin{equation}
T(l,r) =\frac 1{\beta (2\pi)^3}
\int d^4x\is dk  e^{-tk^2} \langle
L(l) R(r) \rangle_k \label{ex1}
\end{equation}
with some functions $r(x)$ and $l(x)$. For example, taking $l=l_2$
and $r=tr_2$ the expression (\ref{ex1}) reproduces the first term
$(tE)$ in (\ref{expk3}). Let us expand $r(x)$ and $l(x)$ in the
Fourier integrals
\begin{eqnarray}
&&r(x) = \frac 1{\beta^{1/2}(2\pi)^{3/2}}\, \is dq\,
\tilde r(q) e^{\imath qx},\nonumber\\
&&l(x)= \frac 1{\beta^{1/2}(2\pi)^{3/2}}\, \is dq'\,
\tilde l(q') e^{\imath q'x} .
\label{Fourier}
\end{eqnarray}
Then
\begin{equation}
\langle L(l)R(r) \rangle_k =\frac 1{\beta (2\pi)^3} \is dq\, \is
dq'\, \tilde r(q)\tilde l(q') e^{\imath (q+q')x}e^ {-\frac \imath
2 k\wedge (q-q') -\frac \imath 2 (q'-k)\wedge (q+k)}
\,,\label{fourLR}
\end{equation}
where
\begin{equation}
k\wedge q\equiv \theta^{\mu\nu} k_\mu q_\nu \,.\label{dwedge}
\end{equation}
Next we substitute (\ref{fourLR}) in (\ref{ex1})
and integrate over $x$ and $q'$ to obtain
\begin{equation}
T(l,r)=  \frac 1{\beta (2\pi)^3}\is dk\, \is dq e^{-tk^2}
\tilde l(-q)\, \tilde r(q) e^{-\imath k\wedge q}\,.\label{ex2}
\end{equation}
In our case $k\wedge q=\theta^{4i}(k_4q_i-k_iq_4)$.

Next we study the integral over $k$. The sum over $k_4$ is treated
with the help of the Poisson formula
\begin{equation}\label{seventeen}
\sum_{n\in \mathbb{Z}}f(2\pi n)=\frac{1}{2 \pi}\sum_{n\in\mathbb{Z}}
\int_{-\infty}^{\infty}f(p)e^{-\imath n p} dp.
\end{equation}
We apply this formula to the sum
\begin{equation}
\sum_{k_4} \exp (-tk_4^2 -\imath \theta^{4j}k_4 q_j) \,,\label{sumk4}
\end{equation}
which corresponds to the choice
\begin{equation}
f(p)=\exp \left( -\frac{tp^2}{\beta^2} - \frac{\imath
\theta^{4j}q_jp}\beta \right)
\end{equation}
in (\ref{seventeen}). The sum (\ref{sumk4}) is transformed to
(after changing the integration variable $y=p/\beta$)
\begin{eqnarray}
&&\frac \beta{2\pi} \sum_{n\in \mathbb{Z}} \int_{-\infty}^\infty dy
\exp (-ty^2 - \imath y (\theta^{4j}q_j +\beta n)),\nonumber\\
&&=\frac \beta{2\pi} \sum_{n\in \mathbb{Z}} \sqrt{\frac {\pi}t}
\exp \left( -\frac{ (\theta^{4j}q_j +\beta n)^2}{4t} \right).
\label{trsum}
\end{eqnarray}
The integral over $k_j$ is Gaussian and can be easily performed.
We arrive at
\begin{equation}
T(l,r)=\frac 1{(4\pi t)^2} \is dq \sum_n
\exp \left( -\frac{ |\theta |^2 q_4^2 + (\theta^{4j}q_j
+\beta n)^2}{4t} \right)
h(q),\label{Ttra}
\end{equation}
where
\begin{equation}
h(q)\equiv \tilde l (-q)\, \tilde r(q),
\qquad |\theta |^2\equiv \theta^{4j}\theta^{4j} .\label{hq}
\end{equation}

In eq.\ (\ref{Ttra}) one can still put $|\theta|=0$ thus returning
to the commutative case. The limit $|\theta|\to 0$ does not
commute however with taking the asymptotic $t\to 0$. From now on
we assume $|\theta|\ne 0$. Obviously, all terms in the sum over
$q_4$ are exponentially small as $t\to +0$ except for $q_4=0$.
\begin{equation}
T(l,r)=\frac 1{(4\pi t)^2} \int d^3\bar q \sum_n \exp \left(
-\frac{(\theta^{4j}q_j +\beta n)^2}{4t} \right) h(0,\bar q)+
\mbox{e.s.t.} \label{Ttra1}
\end{equation}
Let us define two projectors
\begin{equation}
\Pi_{\|}^{ij}=\frac{\theta^{4i}\theta^{4j}}{|\theta|^2},\qquad
\Pi_{\bot}^{ij}=\delta^{ij}-\Pi_{\|}^{ij} \label{proj}
\end{equation}
and split $\bar q$ into the parts which are parallel and
perpendicular to $\theta^{4j}$:  $q_\| = \Pi_\| \bar q$,
$q_\bot=\Pi_\bot \bar q$. Then $d^3\bar q=dq_\| d^2q_\bot$, and
$(\theta^{4j}q_j +\beta n)^2=(|\theta| q_\| +\beta n)^2$. The
asymptotics of the integral over $q_\|$ can be calculated by the
saddle-point method. For each $n$ there is one critical point of
the integrand corresponding to $q_\| = q_\|^{(n)}\equiv -\beta
n/|\theta|$. We expand $h(0,q_\|,q_\bot)$ about these critical
points and take the integral over $q_\|$ to obtain
\begin{equation}
T(l,r) = \frac 1{|\theta|(4\pi t)^{3/2}} \sum_{n\in \mathbb{Z}}
\int d^2q_\bot \left( h(0,q_\|^{(n)},q_\bot) + \frac t{|\theta|^2}
h''(0,q_\|^{(n)},q_\bot) + \dots \right),\label{asT}
\end{equation}
where prime denotes derivative with respect to $q_\|$. This
completes the calculation of small $t$ asymptotics for $T(l,r)$.
Since both $l(x)$ and $r(x)$ are supposed to be smooth, their
Fourier components $\tilde l(q)$ and $\tilde r(q)$ fall off faster
than any power at large momenta, and each term in the asymptotic
expansion is given by a convergent sum and a convergent integral.

The expression (\ref{asT}) is already enough to calculate
mixed (non-planar) contributions
to the heat kernel expansions from the terms inside the brackets
in (\ref{expk3}) which do not contain $k$. (We shall do this in a
moment). Regarding the terms which do contain the momentum $k$,
for our purposes
it is enough to evaluate the power of $t$ appearing in front of
such terms. One can easily trace which modifications appear in the
calculations (\ref{ex1}) - (\ref{asT}) due to the presence of a
polynomial of $k_\mu$. The result is: (i) we still have an expansion
in $t^{1/2}$, (ii) the terms with $k$ do not contribute to the
heat kernel coefficients $a_n$ with $n\le 4$. In other words, the
only relevant mixed heat kernel coefficient is generated by
the first term in the brackets in (\ref{expk3}), and it reads
\begin{equation}
a_3^{\rm mixed}(D)=  \frac 1{|\theta|(4\pi)^{3/2}}\sum_{n\in
\mathbb{Z}} \int d^2q_\bot \tilde l_2 (0,-q_\|^{(n)},-q_\bot)
\tilde r_2(0,q_\|^{(n)},q_\bot),
 \label{a3mix}
\end{equation}
where we substituted the fields appearing in $E$ (see eq.\
(\ref{opD})). Note, that this expression is divergent in the
commutative limit $|\theta|\to 0$. The coefficient $a_3^{\rm
mixed}$ is highly non-local. The structure of (\ref{a3mix}),
especially the sum over $n$, reminds us of the heat kernel
coefficients on NC torus for a rational NC parameter
\cite{Gayral:2006vd}. In this latter case there is a simple
geometric interpretation in terms of periodic projections
\cite{DVCS}. No such interpretation is known for the present case
of $S^1\times \mathbb{R}^3$. However, some similarities can be
found to the works \cite{distime} discussing discretization of the
coordinates which do not commute with a compact coordinate.

\subsection{Renormalization}\label{sec-ren}
Let us return to our particular model (\ref{s1}). First we
summarize the results of the previous subsection and re-express the
heat kernel coefficients $a_n=a_n^{\rm planar}+a_n^{\rm mixed}$ in
terms of the background field $\varphi$ by means of (\ref{choi}):
\begin{eqnarray}
&&a_2(D)=-\frac g{48\pi^2} \int d^4x \varphi^2,\label{a2}\\
&&a_3(D)=-\frac g{6|\theta|(4\pi)^{3/2}}\sum_{n\in \mathbb{Z}}
\int d^2q_\bot \tilde \varphi (0,-q_\|^{(n)},-q_\bot) \tilde
\varphi (0,q_\|^{(n)},q_\bot)\label{a3}\\
&&a_4(D)=\frac 1{16\pi^2} \frac {g^2}{36} \int d^4x
\varphi^4_\star ,\label{a4}
\end{eqnarray}
where tilde is used again to denote the Fourier components.

Next we substitute (\ref{a2}) - (\ref{a4}) in (\ref{near0}) and
(\ref{zeta0}) to the pole part of the regularized effective action
\begin{equation}
W_s^{\rm pole}=-\frac 1{2s} \int d^4x\, \left( \frac g{48\pi^2}
m^2 \varphi^2 +\frac 1{16\pi^2} \frac {g^2}{36} \varphi^4_\star
\right) .\label{Wpole}
\end{equation}
This divergent part of the effective action can be cancelled by an
infinite renormalization of couplings in (\ref{s1})
\begin{equation}
\delta m^2 = \frac {gm^2}{48\pi^2}\, \frac 1s ,\qquad \delta g= \frac
{g^2}{48\pi^2}\, \frac 1s .\label{recoup}
\end{equation}
There can be, of course, also some finite renormalization which we
do not discuss here. Our main physical observation in this
subsection the renormalization (\ref{recoup}) does not depend on
the temperature $1/\beta$.

Here some more comments are in order. It is a very attractive
feature of the zeta function regularization that the non-planar
non-local coefficient $a_3(D)$ does not affect the counterterms.
This coefficient will, however, contribute at some other places,
like the large mass expansion of the one-loop effective action
(see, e.g., \cite{Vassilevich:2003xt}). Moreover, $a_3(D)$ can
lead to troubles in different regularization schemes. For example,
if one uses the proper-time cut-off at some scale $\Lambda$
defining the regularized effective action by
\begin{equation}
W_\Lambda =-\frac 12  \int_{1/\Lambda^2}^\infty \frac{dt}{t}
e^{-tm^2} K(t,D),\label{cutoff}
\end{equation}
the coefficient $a_3$ generates a linear divergence $\propto
\Lambda a_3(D)$, which has no classical counterpart and cannot be
renormalized away in the standard approach. There is a subtraction
scheme (that was used in quantum field theory on curved background
\cite{BD} and in Casimir energy calculations \cite{Bordag:1998vs})
which prescribes to subtract all contributions from several
leading heat kernel coefficients, including $a_3(D)$ in four
dimensions. In the case of two-dimensional scalar theories this
heat kernel subtraction scheme is equivalent to usual
renormalization with the ``no-tadpole'' normalization condition
\cite{Bordag:2002dg}. In the present case the heat kernel
subtraction is, obviously, {\em not} equivalent to the charge and
mass renormalizations given by (\ref{recoup}).

We restricted ourselves to the case of pure space-time
noncommutativity $\theta^{ij}=0$. However, one can try to make an
educated guess on what happens for a generic non-degenerate
$\theta^{\mu\nu}$. By comparing the heat kernel expansion obtained
above with that on NC torus \cite{Gayral:2006vd} and on NC plane
with a non-degenerate $\theta^{\mu\nu}$ \cite{Vassilevich:2005vk}
we can derive (rather unrigorously) the following rule: the
presence of a non-compact NC dimension increases the number of the
first non-trivial field-dependent non-planar (mixed) heat kernel
coefficient by one as compared to the first non-trivial
field-dependent coefficient in the commutative case. Indeed, in
the commutative case the first such coefficient is $a_2$. On the
NC torus \cite{Gayral:2006vd} (no non-compact dimensions) the
first field-dependent mixed heat kernel coefficient is also $a_2$.
For the geometry studied in this paper (one non-compact NC
dimension) this is $a_3$. On an $n$-dimensional NC plane with a
non-degenerate $\theta^{\mu\nu}$ the first coefficient of interest
is $a_{n+2}$ \cite{Vassilevich:2005vk}. We can expect therefore
that the first mixed coefficient on $S^1\times \mathbb{R}^3$ with
a non-degenerate $\theta^{\mu\nu}$ (three compact NC dimensions)
will be $a_5$. Such coefficient does not contribute to one loop
divergence neither in the zeta-function regularization, nor in the
proper-time cut-off scheme. Thus the situation in the generic case
may be expected to better that in the case of a degenerate
$\theta^{\mu\nu}$ discussed above. A similar conclusion has been
made for the Moyal plane in \cite{Gayral:2004cu}.

As we have already mentioned above, the counterterms do not depend
on the temperature. However, if one does the calculations directly
on the zero-temperature manifold $\mathbb{R}^4$, there appears
problems for a degenerate  NC parameter \cite{Gayral:2004cu}.
Perhaps, compactification of one of the NC directions is a proper
way to regularize these problems away.
\section{From imaginary to real time formalism}\label{sec-IR}
The methods which allow to make correspondence between imaginary
and real time formalisms in the case of frequency-dependent
Hamiltonians were suggested in \cite{nlsp} and developed further
in \cite{Fursaev}. Here we briefly outline these methods and
discuss the peculiarities of their application to noncommutative
theories. From now on we work with static background fields,
$\partial_0\varphi = \partial_4 \varphi =0$.
\subsection{Spectral density in the real time formalism}
\label{sec-spden}

Let us consider a Minkowski space counterpart of the action
(\ref{s1}). Our rules for the continuation between Euclidean and
Minkowski signatures read $\partial_4 \to \imath\partial_0$ and
$\theta^{j4}\to -\imath \theta^{j0}$, where $\theta^{j0}$ is real,
and $\theta^{j0}\partial_0$ corresponds to
$\theta^{j4}\partial_4$. We have, therefore, a real NC parameter
in the Moyal product on both Euclidean and Minkowski spaces. These
rules were applied, e.g., in \cite{GoMe}, and they follow also
from the requirement of reflection positivity \cite{Dehne}. As we
shall see below, these rules also ensure consistency between the
expressions for the free energy defined in imaginary and real time
formalisms.

The wave equation for quantum fluctuations $\psi (x)$ over a
static background reads
\begin{equation}
\left(-\partial_0^2 + \partial_j^2 - m^2 -\frac g6
(L(\varphi^2)+R(\varphi^2)+L(\varphi)R(\varphi)) \right) \psi
(x)=0 .\label{waveq}
\end{equation}
The wave operator in (\ref{waveq}) commutes with $\partial_0$.
Consequently, one can look for the solutions $\psi_\omega$ whose
time dependence is described by $\psi_\omega (x) \sim e^{i\omega
x^0}$. They satisfy the equation
\begin{equation}
(P(\omega) + m^2)\psi_\omega = \omega^2 \psi_\omega
\,,\label{weqonom}
\end{equation}
where
\begin{equation}
P(\omega)=-\partial_j^2 +V(\omega),\qquad V(\omega)=\frac g6
(\varphi_+^2+\varphi_-^2+\varphi_+\varphi_-) \label{Pomega}
\end{equation}
and
\begin{equation}
\varphi_{\pm}(x^j)=\varphi \left( x^j\pm \frac 12 \theta^{j0}
\omega \right).\label{phipm}
\end{equation}
Here we used the fact that left (right) Moyal multiplication of a
function of $x^j$ by $\exp(i\omega x^0)$ is equivalent to a shift
of the argument.

From now on we consider the case of positive coupling $g$ only.
Then the potential $V(\omega)$ is non-negative, $V=(g/12)
(\varphi_+^2 +\varphi_-^2 + (\varphi_++\varphi_-)^2)\ge 0$.

To define spectral density for the equation (\ref{weqonom}) we
follow the works \cite{nlsp}. Consider an auxiliary eigenvalue
problem,
\begin{equation}
(P(\lambda)+m^2 ) \psi_{\nu,\lambda} = \nu^2
\psi_{\nu,\lambda}\,.\label{auxsp}
\end{equation}
Obviously, the functions $\psi_{\omega,\omega}$ solve the equation
(\ref{weqonom}).

Our next step differs from that in \cite{nlsp}. Let us restrict
$\lambda$ to $\lambda \le \lambda_0$ for some $\lambda_0$ and put
the system in a three dimensional box with periodic boundary
conditions. Let us suppose that the size of the box is $\gg \theta
\lambda_0$, so that $\varphi_+$ and $\varphi_-$ are localized far
away from the boundaries. In this case, the spectrum of the
regularized problem can be considered as an approximation to the
spectrum of the initial problem for the whole range of $\nu$.
Later we shall remove the box, and the restriction $\lambda \le
\lambda_0$ will become irrelevant. In the box, the spectrum of
$\nu$ in (\ref{auxsp}) becomes discrete, but, for a sufficiently
large box, the spacing is small. The eigenvalues
$\nu_N^2(\lambda)$ depend smoothly on $\lambda$ not greater than
$\lambda_0$, and we can define the density of states as
\begin{equation}
\frac {dn(\nu,\lambda)}{d(\nu^2)}=\frac 1{2\nu} \, \frac
{dn(\nu,\lambda)}{d\nu}= \sum_N \delta (\nu^2 -
\nu^2_N(\lambda)),\label{nudens}
\end{equation}
which can be used to calculate spectral functions of $\tilde
P(\lambda)$, where tilde reminds us that we are working with a
finite-volume problem. For example,
\begin{equation}
\widetilde{\rm Tr} (e^{-t(\tilde P(\lambda) +m^2}) = \int_m^\infty
\frac{dn(\nu,\lambda)}{d\nu} e^{-t\nu^2}\, d\nu \,.\label{htrbox}
\end{equation}
Here $\widetilde{\rm Tr}$ denotes the $L_2$ trace in the box. The
potential $V$ is non-negative. Consequently, there are no
eigenvalues below $m$.

The eigenvalues $\omega_N^2$ of the initial problem
(\ref{weqonom}) in this discretized setting appear when the line
$\nu^2=\lambda^2$ intersects $\nu_N^2(\lambda)$. We can define the
density of the eigenfrequencies $\omega_N^2$ by the formula
\begin{equation}
\frac {dn(\omega)}{d(\omega^2)}=\sum_N \delta (\omega^2 -
\omega_N^2).\label{dodis}
\end{equation}
Next, we would like to relate this density to (\ref{nudens}). This
can be done by calculating derivative of the arguments of the
delta function taken for $\omega = \lambda = \nu$. We obtain
\begin{equation}
\frac {dn(\omega)}{d(\omega^2)}=\frac {d\hat
n(\omega,\omega)}{d(\omega^2)}, \label{dndn}
\end{equation}
where
\begin{equation}
\frac{d\hat n(\nu,\lambda)}{d(\nu^2)}= \sum_N \left(
1-\frac{d(\nu_N^2)}{d(\lambda^2)} \right) \delta (\nu^2 -\nu_N^2)
\,. \label{nhat}
\end{equation}
This density admits an interpretation in terms of the heat kernel
\begin{eqnarray}
&&\widetilde{\rm Tr} \left[ \left( 1 - \frac {1}{2\lambda} \,
\frac{d \tilde P(\lambda)}{d\lambda} \right) e^{-t(\tilde
P(\lambda)+m^2)} \right]= \left( 1 + \frac 1{2\lambda t}\, \frac
{d}{d\lambda} \right) \widetilde{\rm Tr} (e^{-t(\tilde P
(\lambda)+m^2)}) \nonumber\\
&&\quad =\int_m^\infty \frac{d\hat n(\nu,\lambda)}{d\nu}
\,e^{-t\nu^2}\, d\nu \,. \label{hathtr}
\end{eqnarray}

Next we remove the box. Most of the quantities discussed above are
divergent in the infinite volume limit. In order to remove these
divergences we subtract the spectral densities corresponding the
the free operator $\tilde P_0 +m^2$ with $\tilde
P_0=-\partial_j^2$ (not to be confused with $\tilde P(0)$). Then
we perform the infinite volume limit. The limits of subtracted
densities $dn(\omega)/d\omega$, $dn(\nu,\lambda)/d\lambda$ and
$d\hat n(\nu,\lambda)/d\lambda$ will be denoted by $\rho (\omega)$,
$\rho(\nu,\lambda)$ and $\varrho(\nu,\lambda)$, respectively. The
following relation holds in this limit:
\begin{equation}
{\rm Tr}_3 \left( e^{-t(P(\lambda)+m^2)}\right)_{\rm sub} =
\int_m^\infty d\omega \, \rho (\omega;\lambda) \,e^{-t\omega^2}
,\label{htrdo}
\end{equation}
where ${\rm Tr}_3$ is the $L_2$ trace on $\mathbb{R}_3$ and
\begin{equation}
{\rm Tr}_3 \left( e^{-t(P(\lambda)+m^2)} \right)_{\rm sub} \equiv
{\rm Tr}_3 \left( e^{-t(P(\lambda)+m^2)} - e^{-t(-\partial_j^2
+m^2)} \right).\label{def3sub}
\end{equation}
We also have the relation
\begin{equation}
\left( 1 + \frac 1{2\lambda t}\, \frac d{d\lambda} \right) {\rm
Tr}_3 \left( e^{-t(P(\lambda)+m^2)} \right)_{\rm sub}=
\int_m^\infty \varrho (\nu;\lambda)\, e^{-t\nu^2} \, d\nu \,,
\label{1233}
\end{equation}
which, together with (\ref{htrdo}), yields
\begin{equation}
\varrho (\omega;\lambda) = \rho (\omega;\lambda) + \frac
{\omega}{\lambda} \int_m^\omega \partial_\lambda \rho
(\sigma;\lambda)\, d\sigma\,. \label{twodens}
\end{equation}
To derive this formula one has to integrate by parts. Vanishing of
the boundary terms is established by using the same arguments as
in \cite{nlsp}. An infinite volume counterpart of (\ref{dndn})
reads
\begin{equation}
\rho (\omega)=\varrho (\omega;\omega).\label{rvr}
\end{equation}

An independent calculation of the spectral densities is a very
hard problem. We shall view the equation (\ref{htrdo}) as a
definition of the subtracted spectral density $\rho (\nu;\lambda)$
through the heat kernel (an explicit formula involves the inverse
Laplace transform). The other spectral densities $\varrho
(\nu;\lambda)$ and $\rho (\omega)$ are then defined through
(\ref{twodens}) and (\ref{rvr}).

Relations similar to (\ref{htrdo}), (\ref{1233}), (\ref{twodens})
and (\ref{rvr}) were originally obtained in \cite{nlsp} for a
different class of frequency dependent operators and by a somewhat
different method.

\subsection{Wick rotation}\label{sec-Wr}
In this section we show that the Wick rotation of the free energy
$F$ defined through the Euclidean effective action coincides with
the canonical free energy $F_C$. The methods we use are borrowed
from \cite{nlsp}, but there are some subtle points related to
specific features of NC theories. By definition,
\begin{equation}
W(\beta)=\beta (F(\beta) + \mathcal{E}),\label{WFE}
\end{equation}
where $\mathcal{E}$ is the energy of vacuum fluctuations.

Our renormalization prescription (\ref{recoup}) ia equivalent to
the (minimal) subtraction of the pole term (\ref{Wpole}) in
(\ref{near0}). Therefore, the renormalized one-loop effective
action reads
\begin{equation}
W=-\frac 12 \, \frac {d}{ds}\vert_{s=0} \left( \tilde \mu^{2s}
\zeta (s,D+m^2) \right),\label{Wren}
\end{equation}
where $\tilde\mu^2:=\mu^2 e^{-\gamma_E}$. On a static background one
can separate the frequency sum from the $L_2(\mathbb{R}^3)$ trace and
rewrite the zeta function as
\begin{eqnarray}
\zeta (s,D+m^2)&=&\sum_l {\rm Tr}_3 \left( (\omega_l^2 +m^2 +P(\omega_l)
)^{-s} - (\omega_l^2 + m^2 -\partial_j^2)^{-s} \right)\nonumber\\
&=&\sum_l \int_m^\infty d\nu\, \rho_E(\nu;\omega_l) (\omega_l^2
+\nu^2)^{-s}. \label{1804}
\end{eqnarray}
$\omega_l=2\pi l/\beta$. The spectral density $\rho_E$ is defined
for the Euclidean space NC parameter $\theta^{j4}$. It is related
to the real-time spectral density by the formula
\begin{equation}
\rho_E(\nu;\omega|\theta^{j4})=\rho (\nu;\imath \omega
|-i\theta^{j4})
\end{equation}
according to the rules which we have discussed at the beginning of
sec.\ \ref{sec-spden}. We have already mentioned that the Wick
rotation leaves the combination $\omega\theta$ and the potential
$V$ invariant. Therefore, both densities {\it coincide} as
functions of their arguments $\nu$ and $\omega$. However, we shall
keep the subscript $E$ to avoid confusion, but shall drop $\theta$
from the notations for the sake of brevity. Next we use the
formula
\begin{equation}
\sum_l f(\omega_l)=\frac \beta{4\pi \imath} \oint_C \cot \left(
\frac {\beta z}2 \right) f(z) \, dz \label{1830}
\end{equation}
with the contour $C$ consisting of two parts, $C_+$ running from
$\imath \epsilon +\infty$ to $\imath\epsilon -\infty$ and $C_-$
running from $-\imath \epsilon -\infty$ to $-\imath\epsilon
+\infty$, to rewrite the frequency sum as an integral. Then, by
using the symmetry of the integrand with respect to reflections of
$z$ we replace the integral over $C$ by twice the integral over
$C_+$ alone. Finally, we apply the identity
\begin{equation}
\cot \left( \frac {\beta z}2 \right) = \frac 2\beta \, \frac d{dz}
\ln (1-e^{\imath \beta z}) -\imath \label{1833}
\end{equation}
to arrive at the result
\begin{equation}
\zeta (s,D+m^2)=\beta \zeta_0(s,D+m^2)+\zeta_T (s,D+m^2),\label{zzz}
\end{equation}
where
\begin{eqnarray}
&&\zeta_0(s,D+m^2) =\frac 1{\pi} \int\limits_m^\infty d\nu
\int\limits_0^\infty
\rho_E (\nu;z) (\nu^2+z^2)^{-s} dz \,,\label{z0}\\
&&\zeta_T(s,D+m^2) =\frac 1{\pi\imath} \int\limits_m^\infty d\nu
\oint\limits_{C_+} dz\, \left[ \frac d{dz} \ln (1-e^{\imath \beta z})
\right] \rho_E (\nu;z) (\nu^2+z^2)^{-s}\,.\label{zT}
\end{eqnarray}
In commutative theories \cite{nlsp}, the function $\zeta_T$, which
vanishes at zero temperature, represents the purely thermal part,
while $\zeta_0$ is responsible for the vacuum energy. In
space-time NC theories there is no good definition of the
canonical Hamiltonian and of the energy. Therefore, we have no
other choice than to accept the same identities as in the
commutative case, namely
\begin{eqnarray}
&&F(\beta)=-\frac 1{2\beta}   \frac {d}{ds}\vert_{s=0}
\tilde\mu^{2s}\zeta_T(s,D+m^2),\label{FEzT}\\
&&\mathcal{E}=-\frac 1{2}   \frac {d}{ds}\vert_{s=0}
\tilde\mu^{2s}\zeta_0(s,D+m^2).\label{Ez0}
\end{eqnarray}
Actually, the definition of $\mathcal{E}$ is a rather natural one
since it coincides with the renormalized Euclidean one-loop
effective action on $\mathbb{R}^4$. However, as we have already
mentioned in sec.\ \ref{sec-ren} the renormalization in NC
theories depends crucially on the number of compact dimension.
Therefore, if one does the renormalization directly in
$\mathbb{R}^4$, one may need the counterterms which differ from
(\ref{recoup}) obtained on $S^1\times \mathbb{R}^3$.

From now on we concentrate exclusively on $F_T(\beta)$ and
$\zeta_T$. We integrate by parts over $z$ to obtain
\begin{equation}
\zeta_T(s)=-\frac 1{\pi\imath} \int\limits_m^\infty d\nu
\oint\limits_{C_+} dz\, \ln (1-e^{\imath \beta z}) \left[
\frac{\partial_z \rho_E (\nu;z)}{(z^2+\nu^2)^s} -
\frac{2zs\,\rho_E(\nu;z)}{(z^2+\nu^2)^{s+1}} \right]. \label{1920}
\end{equation}
To ensure the absence of the boundary terms we have to deform the
contour $C_+$ by moving its' ends up in the complex plane, so that
$e^{\imath \beta z}$ provides the necessary damping of the
integrand. We discuss the conditions on $\rho_E$ which make such
deformations of the contour legitimate below. The integration by
parts over $\nu$ in the first term in the square brackets in
(\ref{1920}) yields
\begin{equation}
\zeta_T(s)=\frac s{\pi\imath} \int\limits_m^\infty d\nu
\oint\limits_{C_+} dz\, \ln (1-e^{\imath \beta z}) \frac
{2z}{(z^2+\nu^2)^{s+1}} \varrho_E (\nu;z)\,, \label{1924}
\end{equation}
where
\begin{equation}
\varrho_E (\nu;z)=\rho_E(\nu;z) -\frac {\nu}z \int_m^\nu
\partial_z \rho_E (\sigma;z)d\sigma\,. \label{1928}
\end{equation}
The right hand side of (\ref{1924}) is proportional to $s$. To
estimate the derivative $\partial_s $ at $s=0$ in (\ref{FEzT}) one
can put $s=0$ in the rest of the expression and use the Cauchy
theorem after closing the contour in the upper part of the complex
plane. The result is then given by the residue at $z=i\nu$. Next
we make the Wick rotation of the NC parameter, so that
$\rho_E(\sigma;i\nu)$ becomes $\rho (\sigma;\nu)$, and
$\varrho_E(\nu;i\nu)$ becomes $\rho(\nu;\nu)=\rho(\nu)$ (cf. eqs.
(\ref{twodens}) and (\ref{rvr})). Consequently, the Euclidean free
energy is given by the equation
\begin{equation}
F(\beta)=\frac 1\beta \int\limits_m^\infty d\nu \,\rho(\nu) \ln
(1-e^{-\beta\nu})\,, \label{FE}
\end{equation}
which coincides with the canonical definition of the free energy
$F_C$.

The equality $F=F_C$ is the main result of this section. To derive
it we integrated by parts and deformed the contour $C_+$. The
integration by part over $\nu$ is a safe operation, since for any
fixed $z$  the spectral density $\rho_E(\nu,z)$ corresponds to the
Laplace operator with a smooth potential. The absence of the
boundary terms can be then demonstrated by standard arguments
\cite{nlsp} based on the heat kernel expansion. The deformations
of the contour are more tricky. To justify this procedure and
application of the Cauchy theorem one has to assume that
$\rho_E(\nu,z)$ can be analytically continued to the upper
half-plane as an entire function of $z$. A rigorous proof of this
assumption is hardly possible even in more tractable cases of
stationary commutative space-times \cite{nlsp}. We may argue,
however, that this assumption is plausible. Consider pure
imaginary values of $z=i\kappa$. All deformations of the contour
are done before the Wick rotation of the NC parameter $\theta$.
Therefore, $\varphi_\pm$ becomes complex, and
$\varphi_+=\varphi_-^*$. The potential $V(i\kappa)$ remains real
and positive. The background field $\varphi$ is assumed to fall
off faster than any power of the coordinates in real directions to
ensure the existence of the heat kernel expansion. Such fields
typically grow in imaginary directions (one can consider $\varphi
\sim e^{-cx^2}$ as an example). Large positive potentials tend to
diminish the spectral density thus preventing it from the blow-up
behavior. It seems therefore, that the spectral density in our
case should not behave worse than the spectral density in the
commutative case. Another argument in favor of our assumption
will be given at the end of the next section.

The free energy (\ref{FE}) is expressed through a thermal
distribution over the eigenfrequencies. In the absence of a
well-defined Hamiltonian it is not guaranteed that this is the
same as a thermal distribution of one-particle energies. This is a
known problem of space-time NC theories which is beyond the scope
of this paper.
\section{High temperature asymptotics}\label{sec-hT}
As in the previous section we rewrite the regularized one-loop
effective action (\ref{zereg}) on a static background in the form
\begin{equation}
W_s=-\frac 12 \mu^{2s} \Gamma(s) \sum_{\omega} {\rm Tr}_3\,
(\omega^2 + P(\omega) + m^2)^{-s}_{\rm sub} ,\label{Wsom}
\end{equation}
where the sum over the Matsubara frequencies is separated from the
trace over the $L_2$ functions on $\mathbb{R}^3$. As usual, we
subtracted the free space contributions corresponding to
$\varphi=0$ in $P(\omega)$ (which is indicated by the subscript
"sub" in (\ref{Wsom})). We remind that $\omega = 2\pi l/\beta$,
$l\in \mathbb{Z}$. To evaluate the high temperature (small
$\beta$) asymptotics of $W_s$ we split the sum in (\ref{Wsom}) in
two parts,
\begin{equation}
W_s=W_s^{l=0}+W_s^{l\ne 0}, \label{Wsplit}
\end{equation}
which will be treated separately.

We start with $W_s^{l=0}$ which reads
\begin{equation}
W_s^{l=0}=-\frac 12 \mu^{2s} \Gamma (s)\, {\rm Tr}_3\,
(P(0)+m^2)^{-s}_{\rm sub}= -\frac 12 \mu^{2s} \Gamma (s)\, \zeta
(s,P(0)+m^2). \label{Ws01}
\end{equation}
(The subtraction of free space contributions is included in our
definition of the zeta function, cf. (\ref{defzeta})). For each
given $\omega$ the operator $P(\omega)$ is a three-dimensional
Laplace operator with a scalar potential. All effects of the
noncommutativity are encoded in the form of this potential.
Therefore, as for all Laplace type operators on $\mathbb{R}^3$,
the zeta function in (\ref{Ws01}) vanishes at $s=0$ making
$W_s^{l=0}$ finite. We can immediately take the limit $s\to 0$ to
obtain the renormalized expression
\begin{equation}
W^{l=0}=-\frac 12 \zeta'(P(0)+m^2).\label{Wl0ren}
\end{equation}

In the rest of the frequency sum we first use an integral
representation for the zeta function
\begin{equation}
W_s^{l\ne 0}=-\frac 12 \mu^{2s} \sum_{\omega \ne 0} \int_0^\infty
dt\, t^{s-1} {\rm Tr}_3 \left( e^{-t(\omega^2 +m^2 +P(\omega))}
\right)_{\rm sub} \label{Wlne1}
\end{equation}
and then use a trick similar to the one employed in the previous
section. Namely, we replace the operator in the exponential on the
right hand side of (\ref{Wlne1}) by $\omega^2 +m^2 + P(\lambda)$,
expand each of the terms under the frequency sum in asymptotic
series at $\omega\to \infty$ keeping $\lambda$ fixed, and then put
$\lambda =\omega$. The result of this procedure reads
\begin{eqnarray}
W_s^{l\ne 0}&=&-\frac 12 \mu^{2s} \sum_{\omega\ne 0}
\sum_{n=2}^\infty \int_0^\infty dt\, t^{s-1}\, t^{\frac{n-3}2}
\, e^{-t\omega^2} a_n(P(\omega)+m^2)_{\rm sub} \nonumber\\
&=&-\frac 12 \mu^{2s} \sum_{\omega\ne 0} \sum_{n=2}^\infty
|\omega|^{3-n-2s} \Gamma \left( \frac{n-3}2 +s\right)
a_n(P(\omega)+m^2)_{\rm sub}.\label{Wl2}
\end{eqnarray}
Some comments are in order. Here we used again the fact that
$\omega^2 + m^2 + P(\lambda)$ for a fixed $\omega$ is just a usual
Laplace type operator in three dimensions. The large $\omega$
expansion of the heat trace in (\ref{Wlne1}) is therefore standard
and, as well as the usual large mass expansion is defined by the
heat kernel coefficients (see, e.g., \cite{Vassilevich:2003xt}).
On a manifold without boundary an asymptotic expansion
(\ref{asymptotex}) with the replacement $D\to P+m^2$ exists, and
only even numbers $n$ appear. The coefficient $a_0$ vanishes due
to the subtraction, so that the sum in (\ref{Wl2}) starts with
$n=2$.

Now, we have to study the behavior of $a_n(P(\omega)+m^2)_{\rm
sub}$ at large $\omega$. These heat kernel coefficients are
integrals over $\mathbb{R}^3$ of polynomials constructed from the
potential $V(\omega)$ and its derivatives. We can present them as
\begin{equation}
a_n(P(\omega)+m^2)_{\rm sub}=a_n(P+m^2)_{\rm sub}^{\rm planar}+
a_n(P(\omega)+m^2)^{\rm mixed},\label{anpmix}
\end{equation}
where the first (planar) contribution contains all terms which are
the products of either $\varphi_+$ and its derivatives only, or of
$\varphi_-$ and its derivatives only (but not the products of
$\varphi_+$ and $\varphi_-$). The rest is collected in the second
(mixed) contribution. Obviously, no subtraction for the mixed heat
kernel coefficient is needed. Due to the translation invariance of
the integral over $\mathbb{R}^3$, the planar coefficient does not
depend on $\omega$. E.g., $\int d^3x \varphi_+^2=\int d^3x
\varphi^2$. Therefore, we drop $\omega$ from the notation.

First, let us consider the mixed contributions to (\ref{Wl2}). We
assumed that the background field $\varphi$ belongs to $C^\infty
(S^1\times \mathbb{R}^3)$. Therefore, it should vanish
exponentially fast at large distances. Since each term in
$a_n(P(\omega)+m^2)^{\rm mixed}$ contains a product of at least
one $\varphi_+$ with at least one $\varphi_-$, it should be of
order $C_2 e^{-C_1 |\omega \theta|}$ for large $\omega$, where
$C_2$ and $C_1$ are some constants. $C_1$ is positive and
characterizes the fall-off of $\varphi$ at large distances. $C_2$
depends on $n$, on the amplitude of $\varphi$, and on the
functional form of $a_n$. Up to an inessential overall constant
the contribution of a mixed coefficient to (\ref{Wl2}) can be
estimated as
\begin{equation}
\sim \sum_{\omega\ne 0} |\omega|^{3-n} e^{-C_1|\omega \theta|}
\sim \sum_{l=1}^\infty \beta^{n-3} l^{3-n} \exp \left( -\frac{2\pi
C_1 l|\theta| }{\beta} \right) \label{Wmix1}
\end{equation}
(this sum is obviously convergent, so that one can remove the
regularization parameter). If $\beta$ is small enough, namely
$\beta\ll C_1 |\theta|$, all terms in the sum (\ref{Wmix1}) are
strongly suppressed, and the value of the sum can be well
approximated by the first term
\begin{equation}
\sim \beta^{n-3} \exp \left( -\frac{2\pi C_1 |\theta| }{\beta}
\right). \label{Wmix2}
\end{equation}
We conclude that the contributions of the mixed terms are
exponentially small and can be neglected\footnote{If one imposes a
stronger restriction on the background requiring that $\varphi$ is
of compact support, then the mixed terms vanish identically above
certain temperature.} in the high temperature expansion of the
effective action.

Since the planar heat kernel coefficients do not depend on
$\omega$ we are ready to evaluate their contribution to
(\ref{Wl2}) by using precisely the same procedure as in Dowker and
Kennedy \cite{DowKen}.
\begin{eqnarray}
&&W_s^{l\ne 0}= -\mu^{2s} \sum_{n=2}^\infty \sum_{l=1}^\infty
\Gamma \left( \frac {n-3}2 +s \right)\, l^{3-n-2s} a_n(P+m^2)_{\rm
sub}^{\rm planar} \left( \frac \beta{2\pi} \right)^{n-3+2s}
\nonumber\\
&&=-\mu^{2s} \sum_{n=2}^\infty \Gamma \left( \frac {n-3}2 +s
\right)\, \zeta_R(2s+n-3) a_n(P+m^2)_{\rm sub}^{\rm planar} \left(
\frac \beta{2\pi} \right)^{n-3+2s}\label{0129}
\end{eqnarray}
with $\zeta_R$ being the Riemann zeta function. We remind that the
index $n$ in (\ref{0129}) is even. The only divergence in
(\ref{0129}) is a pole in $\zeta_R$ for $n=4$. The corresponding
term near $s=0$ reads
\begin{equation}
\frac 12 a_4(P+m^2)_{\rm sub}^{\rm planar} \frac
{\beta}{(4\pi)^{1/2}} \left[ -\frac 1s -\gamma_E -2 \ln \left(
\frac{\mu \beta}{4\pi} \right) \right]. \label{Wa4}
\end{equation}
On static backgrounds there is a useful formula which relates
planar heat kernel coefficients of $D$ and $P$,
\begin{equation}
a_n(D+m^2)_{\rm sub}^{\rm planar}=\frac
{\beta}{(4\pi)^{1/2}}a_n(P+m^2)_{\rm sub}^{\rm planar}.
\end{equation}
This formula follows from the analysis of the planar heat kernel
coefficients presented in sec.\ \ref{sec-hk} and general formulae
for the heat kernel expansion of Laplace type operators
\cite{KirstenBook,NewGilkey,Vassilevich:2003xt}. The coefficient
$\beta$ appears due to the integration of a constant function over
the Euclidean time, and $(4\pi)^{1/2}$ comes from different
prefactors in the heat kernel coefficients in 3 and 4 dimensions.
In particular, $a_4(D+m^2)^{\rm planar}_{\rm sub}=-m^2a_2(D) +
a_4(D)= \beta /(4\pi)^{1/2} a_4(P+m^2)_{\rm sub}^{\rm planar}$
(let us remind that mixed $a_2(D)$ and $a_4(D)$ vanish). From
(\ref{near0}) and (\ref{zeta0}) we see that the divergence in the
Euclidean effective is reproduced. This divergence is then removed
by the renormalization of couplings (\ref{recoup}). After the
renormalization, we collect all contributions to the effective
action to obtain our final result for the high temperature
expansion of the renormalized effective action
\begin{eqnarray}
&&W=-\frac {{\pi}^{3/2}}{3\beta}\, a_2(P+m^2)^{\rm planar}_{\rm
sub}-\frac 12 \zeta'(P(0)+m^2) \nonumber\\
&&\qquad -\frac 12 a_4(P+m^2)_{\rm sub}^{\rm planar}
\frac{\beta}{(4\pi)^{1/2}} \left[ \gamma_E +2 \ln \left(
\frac{\mu\beta}{4\pi} \right) \right]\label{hTW}\\
&&\qquad -\sum_{n=6}^{\infty} \Gamma \left( \frac{n-3}2 \right)
\zeta_R (n-3) a_n(P+m^2)_{\rm sub}^{\rm planar} \left(\frac
{\beta}{2\pi} \right)^{n-3}.\nonumber
\end{eqnarray}

It is instructive to compare the expansion (\ref{hTW}) to the one
in the commutative case obtained by Dowker and Kennedy
\cite{DowKen} (note, that the normalization of the heat kernel
coefficients used in that paper differs from ours). We see that
the $\zeta'$ term is the same in both cases. The terms
proportional to the heat kernel coefficients for the commutative
case can be obtained from the expansion above by means of the
replacement $a_n(P+m^2)_{\rm sub}^{\rm planar} \to
a_n(P(0)+m^2)_{\rm sub}$. (In both cases subtraction of the free
space contribution means simply deleting the highest power of $m$
in standard analytical expressions \cite{Vassilevich:2003xt}). Let
us write down explicit expressions for a couple of leading heat
kernel coefficients. In the NC case we have
\begin{eqnarray}
&&a_2(P+m^2)_{\rm sub}^{\rm planar}=\frac 1{(4\pi)^{3/2}}
\int d^3x \frac g3 \varphi^2,\nonumber\\
&&a_4(P+m^2)_{\rm sub}^{\rm planar}=\frac 1{(4\pi)^{3/2}}
\int d^3x \left[ \frac {g^2}{36} \varphi^4 + \frac g3 m^2 \varphi^2
\right] .\label{hkNC}
\end{eqnarray}
The coefficients appearing in the commutative case are
\begin{eqnarray}
&&a_2(P(0)+m^2)_{\rm sub}=-\frac 1{(4\pi)^{3/2}}
\int d^3x \frac g2 \varphi^2,\nonumber\\
&&a_4(P(0)+m^2)_{\rm sub}=\frac 1{(4\pi)^{3/2}} \int d^3x \left[
\frac {g^2}{8} \varphi^4 + \frac g2 m^2 \varphi^2 \right]
.\label{hkC}
\end{eqnarray}
In both cases the corresponding heat kernel coefficients differ
only by numerical prefactors in front of the same powers of
$\varphi$.

The high temperature expansion does not depend on $\theta$. In the
limit $\theta\to 0$ (which is a trivial operation) one does not
reproduce the corresponding expansion in the commutative case. The
limits $\beta\to 0$ and $\theta\to 0$ are not interchangeable
because of the condition $\beta\ll C_1|\theta|$ which was imposed
when studying the mixed contributions to the asymptotic expansion.

In space-space NC theories a drastic reduction of the degrees of
freedom in the non-planar sector above certain temperature was
observed in \cite{Fischler:2000fv}.
This may be related in some way to the absence of non-planar contributions
to the high temperature power law asymptotics in space-time NC theories
found above.

One can calculate the high-temperature asymptotics also in the
real-time formalism. The key observation that the non-planar
sector does not contribute remains valid also in this formalism.
The heat kernel expansion in the planar sector has the standard
form (though the values of the heat kernel coefficients differ
from the commutative case). Therefore, one can repeat step by step
the calculations of \cite{nlsp} and obtain an expansion for the
free energy which is consistent with the expansion for the
effective action derived above in the imaginary-time formalism.
This is another argument in favor of the assumptions made in sec.\
\ref{sec-Wr}.
\section{Conclusions}\label{sec-con}
In this paper we considered some basic features of the
finite-temperature NC $\phi^4$ theory in the imaginary-time
formalism. We restricted ourselves to the case of pure space-time
noncommutativity, $\theta^{ij}=0$. We used the zeta function
regularization and the heat kernel methods. Although we found
highly non-local non-planar heat kernel coefficients, such
coefficients do not contribute neither to the one-loop
divergences, nor to the high temperature asymptotics. The theory
can be renormalized at one loop by making charge and mass
renormalizations, as usual. The counterterms do not depend on the
temperature (as long as it is non-zero). We expect that the
renormalization of this theory at zero temperature proceeds
differently. The high temperature expansion of the one-loop
effective action looks similar to the commutative case. The
coefficients of this expansion do not depend on the NC parameter
$\theta$, but again, one has to assume that this parameter is
non-zero.

We have also studied relations between the imaginary and real time
formulations. We found that the Wick rotation of the Euclidean
free energy gives the canonical free energy modulo two
assumptions. One assumptions about the behavior of the spectral
density on the complex plane is of technical nature. Another one
is more fundamental, it concerns the interpretation of the
eigenfrequnecies of perturbations as one-particle energies.

An extension of our results to more general models containing
gauge fields and spinors can be done rather straightforwardly.
Gauge fields are particularly important to make connections to
other approaches \cite{Brandt:2003je,Brandt:2007yg}. Curved space-times will
probably be difficult because of the problems with the heat kernel
expansion. Even in the case of a two-dimensional NC space with a
non-trivial metric the heat kernel coefficients for a (rather
simple) operator are known as power series in the conformal factor
only \cite{DV-2DNC}.

\section*{Acknowledgements}
One of the authors (D.V.V.) is grateful to C.~Dehne for helpful
discussions on noncommutative theories and to D.~Fursaev for
answering endless questions regarding the methods of non-linear
spectral problem. This work was supported in part by FAPESP and by
the grant RNP 2.1.1.1112.

\end{document}